\documentclass{aastex}
\usepackage{spr-astr-addons}
\usepackage{url}

\RequirePackage{color}

\usepackage{latexsym}
\usepackage{graphics}
\usepackage{graphicx}
\usepackage{epsfig}

\begin{document}

\title{ The Nuclear Structure and Associated Electron Capture Rates on Odd-Z Nucleus $^{51}$V in Stellar Matter}
\shorttitle{Stellar capture rates on $^{51}$V}
\author{Muneeb-Ur Rahman\altaffilmark{1}}
\affil{Department of Physics, Islamia College Peshawar, KPK,
Pakistan\\ email: muneebtj@gmail.com}
\author{Jameel-Un Nabi\altaffilmark{2}}
\affil{Faculty of Engineering Sciences, GIK Institute of Engineering
Sciences and Technology, Topi 23640, Swabi, KPK, Pakistan\\ email:
jameel@giki.edu.pk }

\begin{abstract}
The Gamow-Teller strength distribution function, B(GT), for the odd
Z parent $^{51}$V, $N-Z$ =5, up to 30 MeV of excitation energy in
the daughter $^{51}$Ti is calculated in the domain of proton-neutron
Quasiparticle Random Phase Approximation (pn-QRPA) theory. The
pn-QRPA results are compared against other theoretical calculations,
(n, p) and high resolution (d, $^{2}$He) reaction experiments. For
the case of (d, $^{2}$He) reaction the calibration was performed for
$0\leq E_{j} \leq 5.0$ MeV, where the authors stressed that within
this excitation energy range the $\Delta L = 0$ transition strength
can be extracted with high accuracy for $^{51}$V. Within this energy
range the current pn-QRPA total B(GT) strength 0.75 is in good
agreement with the (d, $^{2}$He) experiment's total strength of 0.9
$\pm$ 0.1. The pn-QRPA calculated Gamow-Teller centroid at 4.20 MeV
in daughter $^{51}$Ti is also in good agreement with high resolution
(d, $^{2}$He) experiment which placed the Gamow-Teller centroid at
4.1 $\pm$ 0.4 MeV in daughter $^{51}$Ti. The low energy detailed
Gamow-Teller structure and Gamow-Teller centroid play a sumptuous
role in associated weak decay rates and consequently affect the
stellar dynamics. The stellar weak rates are sensitive to the
location and structure of these low-lying states in daughter
$^{51}$Ti. The calculated electron capture rates on $^{51}$V in
stellar matter are also in good agreement with the large scale shell
model rates.
\end{abstract}
\keywords{Gamow-Teller distribution; electron capture; pn-QRPA;
stellar dynamics, supernovae}

\section{Introduction}
Supernovae are crucial to our very existence and for the dynamical/
morphological development of the universe. They are also at the
nexus of many of the great debates raging among astronomers. These
explosions have enriched the galaxy with oxygen we breathe, the iron
in our blood, the calcium in our bones and teeth, and silicon which
is used in the semiconductor-based electronic industries having
enormous applications in this modern world. The weak interactions
soften and smooth the landscape of star for gravity. These two
forces play a pivotal role in cooking the material inside stellar
kilns and play preeminent role in the make up of galaxies and stars
formation and their physical death (supernova explosion). This is
not only true for the star's energy budget but these forces help
each other in a certain sense to change the composition of the
stellar matter and entropy as well. When the ratio of neutrons to
protons is low, electron capture is a more probable process to
occur. The processes mediated by charge-changing weak interaction
are 1) electron capture:  2) positron capture: 3) $\beta^{-}$ decay
in nucleus: 4) $\beta^{+}$ decay. Gamow-Teller (GT) transitions play
a preeminent role in the collapse of stellar core in the stages
leading to a Type-II supernova. The GT strength distributions from
ground and excited states are used for the calculation of weak decay
rates for the core-collapse supernova dynamics and for probing the
concomitant nucleosynthesis problem \cite{Nabi and Rahman07, Nabi
and Rahman05, Fujita05, Nabi and Rahman11, Zhi11, Sasano12}.

The incarnation of the core-collapse mechanism is the conversion of
a fraction of the gravitational energy into kinetic energy of the
ejecta and internal energy of the inner core of the exploding star
\cite{Bur06}. The core collapse is the most energetic event in the
universe with emitting energy budget of 10$^{46}$ J mostly in the
form of neutrinos. When the temperature of the core becomes equal to
or more than 2.5 $\times$ 10$^{9}$ K, oxygen is set on fire and
silicon burning ensues when the inner core attains the temperature
of 3.5 $\times$ 10$^{9}$ K \cite{Jing-Jing}. At this instant the
nucleosynthesis pushes the elements in the inner core to the iron
group nuclei and when this inner core exceeds the appropriate
Chandrasekhar mass limits it becomes unstable and inaugurates the
implosion of the inner core and consequently announces the death of
the star in an implosion and subsequent catastrophic explosion. The
collapse of the star is very sensitive to the entropy and to the
number of leptons per baryon, $Y_{e}$ \cite{Nabi and
Rahman05,Bethe79}. The neutrinos are considered as main sink of
energy and lepton number until the core density reaches around
$10^{10}g cm^{-3}$. At later stages of the collapse this assumption
is no longer legitimate as electron capture rates increase with
increase in stellar core density. This is primarily due to the
reason that the Fermi energy of the electrons increases considerably
as the stellar density increases by orders of magnitude. Example
giving, for a fixed stellar temperature of 3 $\times$ 10$^{9}$ K,
the Fermi energy of the electrons is 0.00 MeV at a stellar density
of $ 10 gcm^{-3}$. The Fermi energy increases to 0.00 MeV, 0.04 MeV,
2.36 MeV and 23.93 MeV as the stellar core stiffens to a density of
$ 10^{3} gcm^{-3}, 10^{5} gcm^{-3}, 10^{8} gcm^{-3}$ and $10^{11}
gcm^{-3}$, respectively. This increase in Fermi energy of electrons
with increasing density makes the electrons more susceptible for
getting captured and causes the electron capture rates to increase
substantially. For densities greater than $ 10^{11} gcm^{-3}$, the
neutrino mean free paths become shorter and consequently proceed
through all phases of free streaming, diffusion, and trapping.  The
most tightly bound of all the nuclei in the inner core of star is
$^{56}$Fe \cite{Cowan04} and further fusion of the nuclei is
endoergic. The second bottleneck to the synthesis of heavier
elements is the high Z number which poses a high Coulombic barrier
for the charged particles to initiate nuclear reactions at stellar
core temperatures. These impediments to further nucleosynthesis is
rescinded with the help of neutron (which do not feel the Coulomb
barrier) capture processes in the stellar core to form heavier
isotopes beyond iron group nuclei. These processes are further
classified into slow (s-) and rapid (p-) neutron capture processes
depending on the neutron capture time scale $\tau_n$ and beta decay
time scale $\tau_\beta$ for nucleus to endure beta decay. As
discussed earlier, weak interaction and gravity are the mentor that
drives the stellar evolutionary process and its subsequent death in
a cataclysmic explosion. The main weak interaction processes that
play an effective role in the stellar evolution are beta decays,
electron and positron capture processes and neutrinos
emission/capture processes subject to the physical conditions
available for these processes in the stellar core. These weak decay
processes are smitten by Fermi and Gamow-Teller (GT) transitions.
Fermi transitions are straightforward and are important only for
beta decays. Spin-isospin-flip excitations in nuclei at vanishing
momentum transfer are commonly known as GT transitions. For nuclei
with $N > Z$, Fermi transitions are Pauli blocked and the GT
transitions dominate and their calculation is model dependent. These
transitions are ideal probe to test nuclear structure and play
preeminent role in the nucleosynthetic origin of the elements in the
late phases of the stellar life. When the stellar matter is
degenerate then the phase space for the electron in beta decay is
Pauli blocked and electron captures become dominant in the stellar
core producing neutrinos which escape from the surface of the star
and takes away the core energy and entropy as well.  These electron
capture rates and $\beta^{+}$ decay rates are very sensitive to the
distribution of the GT$_{+ }$ strength (in this direction a proton
is changed into a neutron).

$^{51}$V plays a key role in the presupernova stages of evolution of
massive stars, specially prior to the presupernova collapse of the
stellar core \cite{Bet79}. $^{51}$V is an important nucleus which is
thought to play a significant role in neutrino-process productions
and also thought to be a key gama ray source \cite{Haxton04}. In
this paper we present the Gamow-Teller strength distribution
function in electron capture direction (referred to as B(GT$_{+}$)
strength distribution) of $^{51}$V using the pn-QRPA model. The
associated electron capture rates on $^{51}$V is also presented in
stellar environment. We also compare and contrast our calculation
with previous results.

Section 2 describes the theoretical formalism used for the B(GT)
strength and weak decay rates based on pn- QRPA model. Section 3
discusses the pn-QRPA calculated GT strength functions and its
comparison with measurements and previous calculations. The
calculated electron capture rates are presented in Sect. 4. We
finally conclude our findings in Sect. 5.

\section{Model Description}
The pn-QRPA is considered an efficient model to extract the GT
strengths for the ground as well as excited states of the nuclei
present in stellar matter \cite{Nab13}. The transitions from the
excited states contribute effectively to the total electron capture
rate and a microscopic calculation of excited state GT strength
distributions is desirable. The pn-QRPA model is used in the present
work to calculate the B(GT$_{+}$) strength distribution and
associated electron capture rates on odd-Z nucleus $^{51}$V using a
luxurious model space of $7\hbar\omega$. The model is capable of
performing a microscopic calculation of ground \textit{and} excited
states GT strength distribution functions.

The Hamiltonian used in our calculation is of the form
\begin{equation} \label{GrindEQ__1_}
{\rm H}^{{\rm QRPA}} {\rm \; =\; H}^{{\rm sp}} {\rm \; +\; V}^{{\rm
pair}} {\rm \; +\; V}_{{\rm GT}}^{{\rm ph}} {\rm \; +\; V}_{{\rm
GT}}^{{\rm pp}},
\end{equation}
where $H^{sp} $ is the single-particle Hamiltonian, $V^{pair} $ is
the pairing force, $V_{GT}^{ph} $  is the particle-hole (ph) GT
force, and $V_{GT}^{pp} $ is the particle-particle (pp) GT force.
Pairing is treated in the \emph{BCS} approximation, where an assumed
constant pairing force with force strength $G$ ($G_{p}$ and $G_{n}$
for protons and neutrons, respectively) is applied
\begin{eqnarray} \label{GrindEQ__2_}
V^{pair} =-G\sum _{jmj'm'}(-1)^{l+j-m}  c_{jm}^{+} c_{j-m}^{+} \nonumber \\
(-1)^{l'+j'-m'} c_{j'-m'} c_{j'm'},
\end{eqnarray}
here the sum over $m$  and $m'$  is restricted to $m$, ${m'}> 0$,
and $l$ donates the orbital angular momentum.

In the present work, in addition to the well known particle-hole GT
force \cite{Halbleib67,Staudt90,Muto92}
\begin{equation} \label{GrindEQ__3_}
V_{GT}^{ph} =2\chi \sum _{\mu }(-1)^{\mu } Y_{\mu }  Y_{-\mu }^{+},
\end{equation}
with
\begin{equation} \label{GrindEQ__4_}
Y_{\mu } =\sum _{j_{p} m_{p} j_{n} m_{n} }<j_{p} m_{p} |t{}_ {-}
\sigma _{\mu }  |j_{n} m_{n} >c_{j_{p} m_{p} }^{+} c_{j_{n} m_{n} },
\end{equation}
the particle-particle GT force \cite{Soloviev87, Kuzmin88}
\begin{equation} \label{GrindEQ__5_}
V_{GT}^{pp} =-2\kappa \sum _{\mu }(-1)^{\mu }  P_{\mu }^{+} P_{-\mu
},
\end{equation}
with
\begin{eqnarray} \label{GrindEQ__6_}
P_{\mu }^{+} =\sum _{j_{p} m_{p} j_{n} m_{n} }<j_{n} m_{n} |(t{}_{-}
\sigma _{\mu } )^{+} |j_{p} m_{p} > \nonumber \\
(-1)^{l_{n} +j_{n} -m_{n} } c_{j_{p} m_{p} }^{+} c_{j_{n} -m_{n}
}^{+},
\end{eqnarray}
is also taken into account.

The electron capture rate of a transition from the \textit{i}th
state of a parent nucleus (Z, N) to the $jth$ state of the daughter
nucleus $(Z - 1, N + 1)$ is given by

\begin{equation} \label{GrindEQ__7_}
\lambda _{ij}^{ec} \, =\, \ln 2\frac{f_{ij} (T,\rho ,E_{f}
)}{(ft)_{ij} }.
\end{equation}
In Eq.~\ref{GrindEQ__7_}, $f_{ij} $ is the phase space integral. The
$(ft)_{ij} $ of an ordinary $\beta ^{\pm } $ decay from the state
${\left| i \right\rangle} $ of the mother nucleus to the state
${\left| f \right\rangle} $ of the daughter is related to the
reduced transition probability $B_{ij} $ of the nuclear transition
by

\begin{equation} \label{GrindEQ__8_}
(ft)_{ij} \, =\, {D\mathord{\left/ {\vphantom {D B_{ij} .}} \right.
\kern-\nulldelimiterspace} B_{ij}.}
\end{equation}
The D appearing in Eq.~\ref{GrindEQ__8_} is a compound expression of
physical constants,
\begin{equation} \label{GrindEQ__9_}
D\, =\, \frac{2\pi ^{3} \hbar ^{7} \ln 2}{g_{V}^{2} m_{e}^{5} c^{4}
},
\end{equation}
whereas the reduced transition probability of the nuclear transition
is given by
\begin{equation} \label{GrindEQ__10_}
B_{ij} \, =\, B(F)_{ij} \, +\, \left({g_{A} \mathord{\left/
{\vphantom {g_{A}  g_{V} }} \right. \kern-\nulldelimiterspace} g_{V}
} \right)^{2} B(GT)_{ij}.
\end{equation}

The value of D is taken to be 6146 $\pm$ 6 s adopted from Ref.
\cite{Jokinen02} and the ratio of the axial-vector $(g_{A} )$ to the
vector $(g_{V} )$ coupling constant is taken as -1.257. $B(F)_{ij} $
and $B(GT)_{ij} $ are reduced transition probabilities of the Fermi
and GT transitions, respectively. These reduced transition
probabilities of the nuclear transitions are given by
\begin{equation} \label{GrindEQ__11_}
B(F)_{ij} \, =\, \frac{1}{2J_{i} +1} |<j||\sum _{k}t_{\pm }^{k} ||\,
i>|^{2},
\end{equation}
\begin{equation} \label{GrindEQ__12_} B(GT)_{ij} \, =\, \frac{1}{2J_{i} +1}
|<j||\sum _{k}t_{\pm }^{k} \vec{\sigma }^{k} ||\,  i>|^{2}.
\end{equation}

The phase space integrals are
\begin{equation} \label{GrindEQ__13_}
f_{ij} \, =\, \int _{w_{l} }^{\infty }w\sqrt{w^{2} -1}  (w_{m} \,
+\, w)^{2} F(+ Z,w)G_{-} dw,
\end{equation}
where $G_{-} $ is the electron distribution function. $F(+z,w)$ are
the so-called Fermi functions and are calculated according to the
procedure adopted by Ref. \cite{Gove71}. $w$ is the total energy of
the electron including its rest mass, $w_{l} $ is the total capture
threshold energy (rest + kinetic) for electron capture.

The number density of electrons associated with protons and nuclei
is $\rho Y_{e} N_{A} ,$ where $\rho $ is the baryon density, $Y_{e}
$ is the ratio of electron number to the baryon number, and $N_{A} $
is the Avogadro's number.
\begin{equation} \label{GrindEQ__14_}
\rho Y_{e} \, =\, \frac{1}{\pi ^{2} N_{A} } (\frac{m_{e} c}{\hbar }
)^{3} \int _{0}^{\infty }(G_{-}  -G_{+} )p^{2} dp,
\end{equation}
where $p\, =\, (w^{2} -1)^{1/2} $ is the electron or positron
momentum and $G_{+}$ is the positron distribution function.

The total electron capture rate per unit time for a nucleus in
thermal equilibrium at temperature $T$  is then given by
\begin{equation} \label{GrindEQ__15_}
\lambda _{ec} \, =\, \sum _{ij}P_{i} \lambda _{ij}^{ec}.
\end{equation}
Here $P_{i} $ is the probability of occupation of parent excited
states and follows the normal Boltzmann distribution. The summation
over all initial and final states is carried out until satisfactory
convergence in the rate calculations is achieved.

\section{Comparison of Gamow-Teller Strength Distributions for $^{51}$V}
\begin{figure}
\begin{center}
\includegraphics[width=6in,height=5.5in]{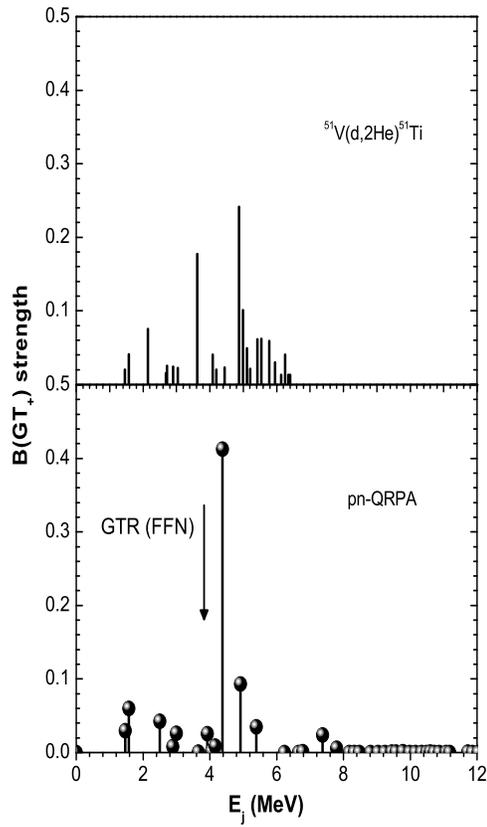}
\caption{Comparison of the B(GT$_{+}$) strength distributions for
$^{51}$V as function of excitation energy in $^{51}$Ti between
pn-QRPA calculation (present work) and (d, $^{2}$He) reaction
\cite{Bäumer03}.  The arrow denotes the position of the centroid of
the GT resonance predicted by FFN \cite{FFN82}.}\label{fig1}
\end{center}
\end{figure}

\begin{figure}
\begin{center}
\includegraphics[width=6in,height=5.5in]{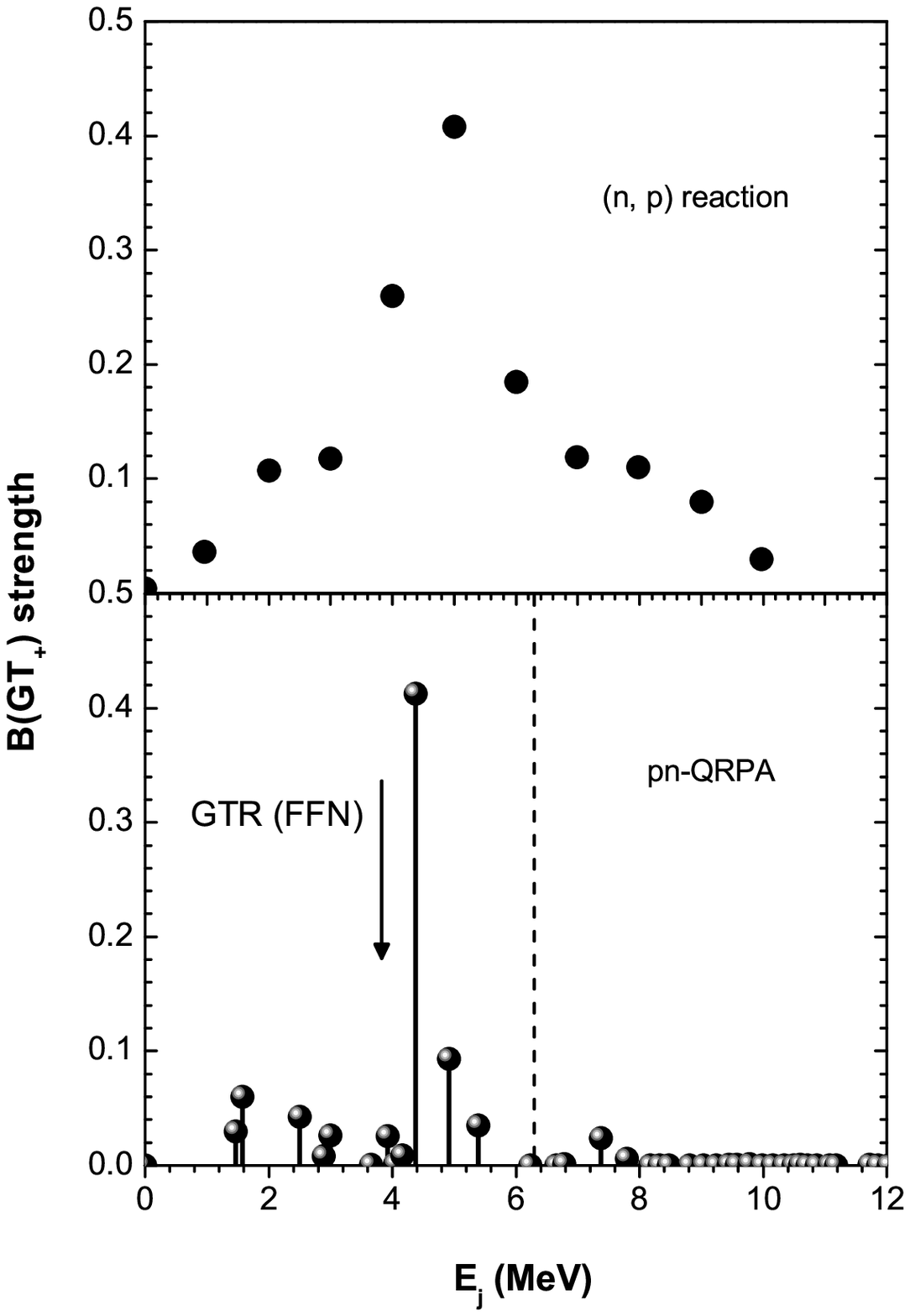}
\caption{Comparison of the B(GT$_{+}$) strength distributions for
$^{51}$V as function of excitation energy in $^{51}$Ti between
pn-QRPA calculation (present work) and (n,p) experiment at TRIUMF
\cite{Alford93}.  The arrow denotes the position of the centroid of
the GT resonance predicted by FFN \cite{FFN82} and the dashed line
represents the maximum energy considered in the (d, $^{2}$He)
reaction experiment \cite{Bäumer03}.}\label{fig2}
\end{center}
\end{figure}

\begin{figure}
\begin{center}
\includegraphics[width=6in,height=5.5in]{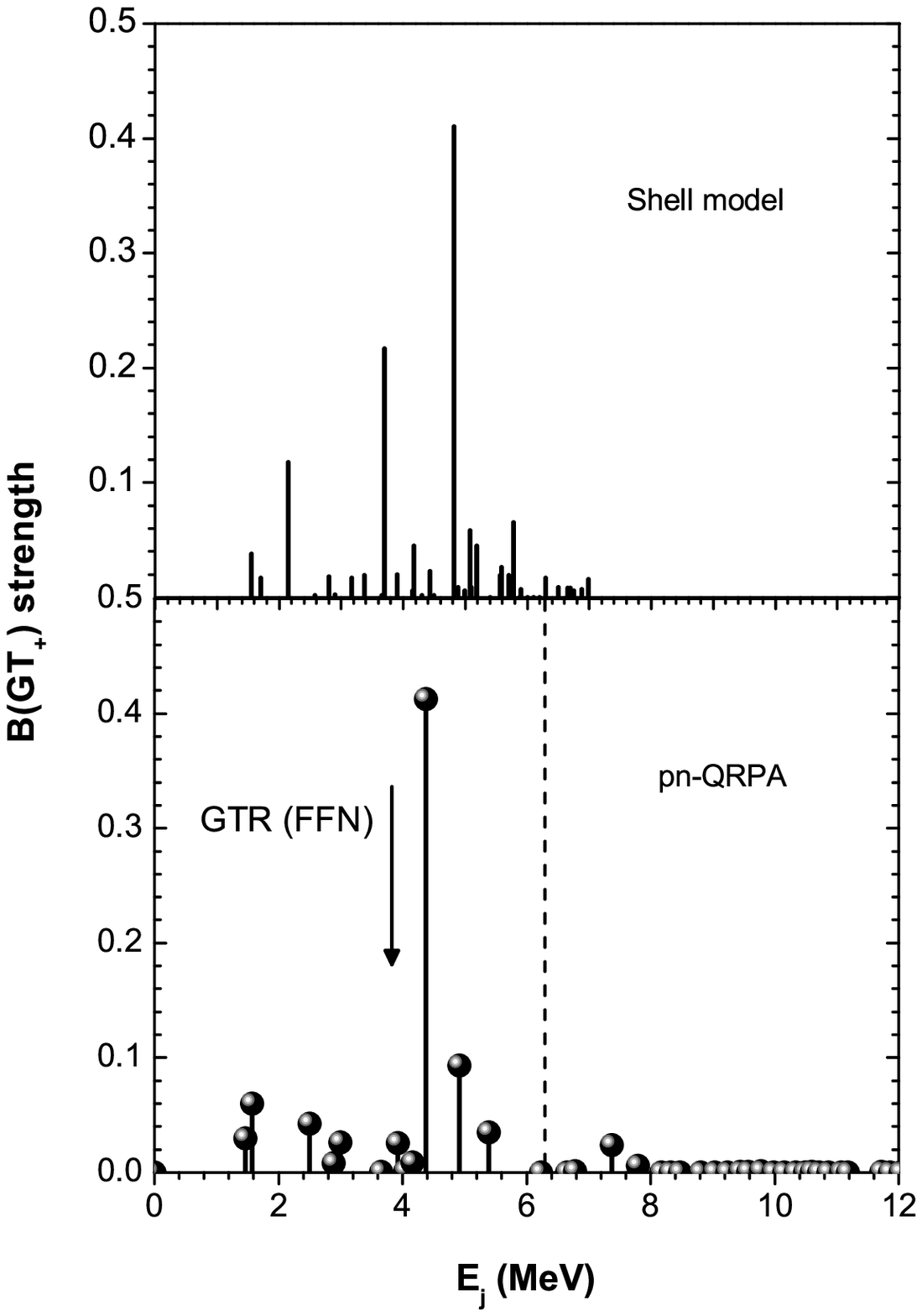}
\caption {Comparison of the B(GT$_{+}$) strength distributions for
$^{51}$V as function of excitation energy in $^{51}$Ti between
pn-QRPA calculation (present work) and shell model \cite{Poves01}
results. The arrow denotes the position of the centroid of the GT
resonance predicted by FFN \cite{FFN82} and the dashed line
represents the maximum energy considered in the (d, $^{2}$He)
reaction experiment \cite{Bäumer03}.}\label{fig3}
\end{center}
\end{figure}

As discussed earlier $^{51}$V is a key nucleus which has a
significant impact on the late stages of evolution of massive stars,
specially prior to collapse \cite{Bet79}. Potentially significant
neutrino-process productions and gama rays sources include many iron
group nuclei including $^{51}$V \cite{Haxton04}. The GT strength
distribution, B(GT), is calculated in the present work for the odd-Z
parent $^{51}$V up to 30 MeV of excitation energy in the daughter
$^{51}$Ti using the pn-QRPA model. The distribution of the GT
strength connecting the ground state of the parent $^{51}$V to the
states in the daughter $^{51}$Ti in the GT$_{+ }$ direction is shown
in Fig.~\ref{fig1} to Fig.~\ref{fig3}. In Figs.~\ref{fig1} and
~\ref{fig2} we compare our data with (d, $^{2}$He) \cite{Bäumer03}
and (n,p) reaction \cite{Alford93} experiments, respectively. The
charge-exchange reactions near $\theta \,=\, 0^{o}$ and at zero
momentum transfer proceed fastidiously through Gamow-Teller
transitions. The ground state of the parent is at $J^{\pi} =
7/2^{-}$, thereby allowing the GT transitions to the states in
daughter with $J^{\pi} = 5/2^{-} ,7/2^{-}$ and $9/2^{-}$ (only
allowed transitions are calculated in this work). Further we also
compare our B(GT$_{+}$) strength distribution calculation with other
theoretical calculations. In Fig.~\ref{fig1} to Fig.~\ref{fig3}, the
pn-QRPA strength is shown up to 12 MeV excitation for the sake of
comparison only. The GT strength of the pn-QRPA is scaled with a
quenching factor of 0.36. Aufderheide and collaborators
\cite{Aufderheide93} used larger model space in their shell model
calculation and renormalized the calculated strength to the measured
GT strength with a factor of 0.31 below 8 MeV.

We first compare our calculation with the (d, $^{2}$He) reaction
experiment. High-resolution studies of GT$_{+}$ transitions are far
more challenging to achieve as compared to studying high-resolution
experiments of GT transitions in the $\beta^{-}$ direction. When the
two protons couple to an $^{1}S_{0}, T = 1$ state, the unbound
diproton system is referred to as $^{2}$He. Experimentally, the
$^{1}S_{0}$ is selected by limiting the relative energy of the
diproton system to 1 MeV. The beam of 171 MeV deuterons was then
targeted on the source (99.75 $\%$ $^{51}$V foils). The (d,
$^{2}$He) reaction experiments are complicated due to the coincident
detection of the two correlated protons in the presence of an
overwhelming background originating from deutron breakup. However,
since the reaction mechanism of (d, $^{2}$He) forces a spin-flip,
the reaction is more selective than the competing (n,p) and (t,
$^{3}$He) reactions. Further details of (d, $^{2}$He) reaction
experiment can be found in Ref. \cite{Bäumer03}. Fig.~\ref{fig1}
compares our B(GT$_{+}$) strength distribution calculation with the
measured (d, $^{2}$He) reaction carried out using the ESN-BBS setup
of the superconducting cyclotron AGOR facility at KVI Groningen
\cite{Bäumer03}. The calibration was performed for $0\leq E_{j} \leq
5.0$  MeV, and there the authors stressed that in this excitation
energy range the transition strength can be extracted with high
accuracy for $^{51}$V. Limiting the excitation energy to 5.0 MeV,
the (d, $^{2}$He) reaction gives total strength of 0.9 $\pm$ 0.1.
The renormalized pn-QRPA total strength 0.79 for $^{51}$V is in good
agreement with the measured strength of Ref. \cite{Bäumer03} for the
same excitation energy range in daughter (additional GT strength at
higher excitation energy above 5 MeV has not been included in this
value). The pn-QRPA result for $^{51}$V is also compared with the
blazing trails of pioneer calculation performed by Fuller, Fowler,
and Newman (FFN) \cite{FFN82} in Fig.~\ref{fig1}. FFN assumed that
almost all the strength is concentrated in a collective state also
referred to as GT resonance (GTR). FFN calculation placed the
centroid of the GTR at 3.83 MeV of the excitation energy (also shown
by a marked arrow in lower panel of Fig.~\ref{fig1}). The pn-QRPA in
present calculation yields GT centroid at 4.20 MeV in daughter
$^{51}$Ti, and is in remarkable agreement with high resolution study
of (d, $^{2}$He) reaction data \cite{Bäumer03}, which placed the
centroid of the GT strength at 4.1 $\pm$ 0.4 MeV.

On the other hand in (n,p) reaction experiments, a primary proton
beam is first targeted on $^{7}$Li. Neutrons are then produced in
the  $^{7}$Li(p,n) reaction with the incoming proton beam of around
200 MeV energy. The resulting neutron beams are then targeted on the
source (in the form of four foils of 99.5$\%$ pure vanadium targets
of varying thickness). For further details of the (n,p) experiment
we refer to \cite{Alford93} and references therein. Fig.~\ref{fig2}
shows the comparison of pn-QRPA calculated B(GT$_{+}$) distribution
with the measured $^{51}$V(n, p) reaction at 198 MeV at TRIUMF
\cite{Alford93}.  The spectroscopic information is available up to
excitation energy 5.22 MeV for the daughter $^{51}$Ti, whereas in
the pn-QRPA calculations discrete levels up to excitation energy of
$E_{j}$ = 8 MeV are seen and beyond this very little strength is
predicted. The (n,p) reaction \cite{Alford93} reports total strength
of 1.2 $\pm$ 0.1 up to 8 MeV for $^{51}$V. Aufderheide et al.
\cite{Aufderheide93} have analyzed the data for $^{51}$V, in the
GT$_{+}$ direction, up to excitation energy 12 MeV in daughter
$^{51}$Ti and extracted total GT strength value of 1.5 $\pm$ 0.2 for
$^{51}$V. From pn-QRPA calculation, one finds little strength above
7 MeV of excitation energy when compared with the measured (n, p)
data of Ref. \cite{Alford93} where reasonable strength is measured
up to 10 MeV in daughter. We again show the placement of GTR
centroid by FFN \cite{FFN82} by a marked arrow in the lower panel of
Fig.~\ref{fig2}. The dashed line at 6.33 MeV in lower panel of
Fig.~\ref{fig2} indicates the maximum excitation energy considered
in the (d, $^{2}$He) reaction experiment \cite{Bäumer03}, and
defines the onset of the flat continuum response in the spectrum. It
will be interesting to snap whether high-lying GT strength exists
above 6.33 MeV in the high resolution study of GT distribution, and
such studies of the GT strength distribution can be done on the
basis of angular distribution and a multipole decomposition
\cite{Bäumer03}.

The shell model calculation employing the KB3G interaction
\cite{Poves01} using a quenching factor of 0.23 for $^{51}$V using
the OXBASH code \cite{Brown86} in truncated model spaces is shown in
Fig.~\ref{fig3}. The shell model B(GT+) values were obtained
adopting the Lanczos method with 100 iterations for each $J^{\pi}$.
The shell model GT strength was normalized to the measured strength
with a factor 0.74.  The pn-QRPA calculated B(GT$_{+}$) strength
distribution seems to be in better accordance with the (d, $^{2}$He)
reaction data \cite{Bäumer03} and shell model calculation
\cite{Poves01} as against the (n,p) reaction experiment
\cite{Alford93}. The shell model result placed the centroid of the
calculated GT distribution at 4.34 MeV in daughter to be compared
with 4.20 MeV calculated by this work. For sake of reference, the
centroid of GTR calculated by FFN \cite{FFN82} is again shown by an
arrow in the lower panel of Fig.~\ref{fig3}. Marked also is the
maximum excitation energy considered in the (d, $^{2}$He) reaction
experiment \cite{Bäumer03} as a dashed line at 6.33 MeV.

Caurier et al. \cite{Caurier99} in their large-scale shell model
calculation employed the monopole corrected version of the KB3
interaction and obtained a total GT strength of 1.4. This total
strength is more in comparison to the reported pn-QRPA total
strength and that measured in (d, $^{2}$He) reaction experiment
\cite{Bäumer03}. The GT centroid in \cite{Caurier99} is placed at a
much higher energy of 5.18 MeV in daughter $^{51}$Ti. This centroid
is around 1 MeV higher in energy than the centroid placements of
pn-QRPA and high resolution (d, $^{2}$He) data \cite{Bäumer03} and
might suppress the weak decay rates of \cite{Caurier99} in the
stellar matter where the density and temperature is comparatively
low.

\section{Electron Capture Rates}
The stellar electron capture rates on $^{51}$V were calculated
within the pn-QRPA formalism using the mass compilation of
\cite{Audi03a, Audi03b} for temperature and density grid as
summarized in Table 1, which additionally lists the chemical
potential. The tabulated chemical potential refers to the electron
kinetic energy and, thus shifted downward by the electron rest mass
relative to the $\mu_e$ values. The three selected temperature
(density) points roughly depict low, medium and high temperature
(density) zones.

\begin{figure}
\begin{center}
\includegraphics[width=6.0in,height=5.5in]{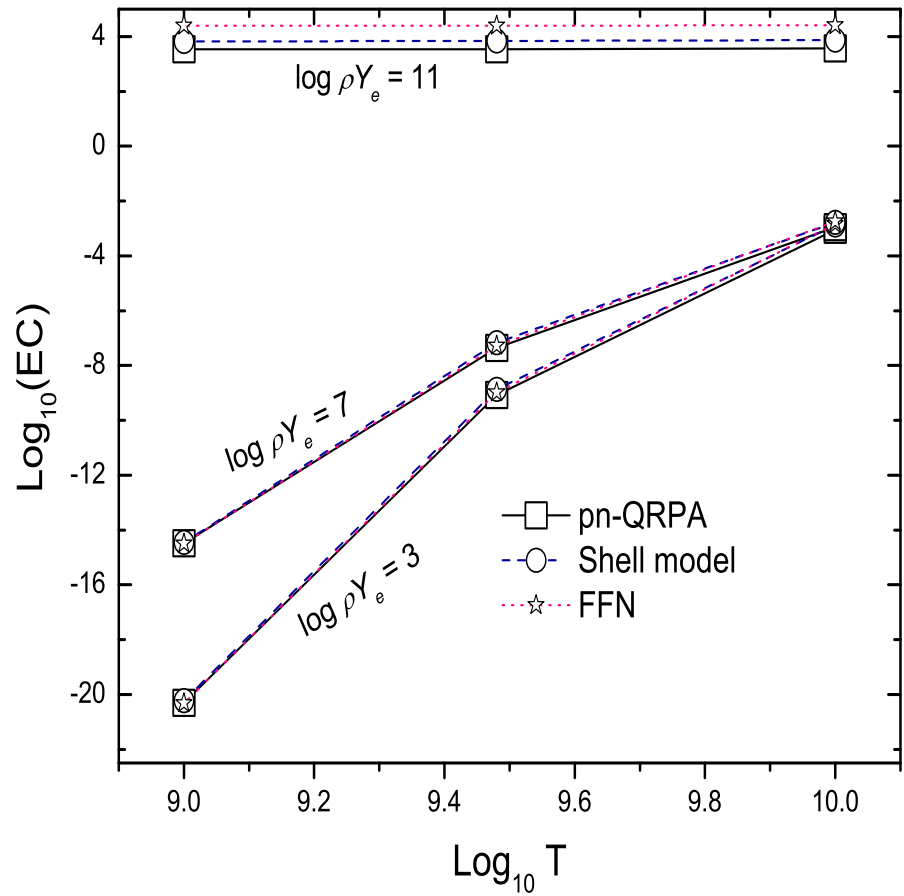} \caption { Comparison of the pn-QRPA,
large scale shell model \cite{Langanke01}, and  FFN \cite{FFN82}
stellar electron capture rate calculations on $^{51}$V as a function
of stellar temperature for selected densities.}\label{fig4}
\end{center}
\end{figure}

The electron capture rates using the pn-QRPA in the present work are
in excellent agreement with the large scale shell model rates
\cite{Langanke01}. The comparison of the two microscopic
calculations is shown in Fig.~\ref{fig4}. It is noted that the
electron captures on $^{51}$V is quite sensitive to the temperature
of the stellar core at low densities. At the first instant in  low
density regions the electrons have to overcome a threshold of nearly
3 MeV and this can be mainly achieved at low ($\rho Y_{e} = 10^{3} g
cm^{-3}$) and intermediate densities ($\rho Y_{e} = 10^{7} g
cm^{-3}$) by the thermal population of excited states in the parent
$^{51}$V. The Fermi energies of the electrons, 0.046 MeV and 1.2
MeV, respectively, for these densities are still smaller than the
Q-value of 2.47 MeV. The population index for the parent increases
with temperature at these low density regime and consequently
electron capture increases with the increase in temperature. As the
stellar core approaches densities prior to presupernova epoch, the
electron capture rates are no more dependent on the stellar
temperature. At this stage the driving force for the electron
captures is the chemical potential which grows much faster than the
nuclear Q-value. Thus, at higher densities sufficient electrons are
available with energies above the Q-value and leads to enhancement
of electron capture rates. It can be seen from Fig.~\ref{fig4} that
for low and medium density domain the shell model rates are in
excellent comparison with the reported rates. Only in high density
range does the shell model rate roughly double the corresponding
pn-QRPA calculated electron capture rate.

Additionally Fig.~\ref{fig4} shows the comparison of pn-QRPA
calculated electron capture rates on $^{51}$V with the calculation
of FFN \cite{FFN82}.  For $^{51}$V, the pn-QRPA electron capture
rates are in good agreement with FFN rates \cite{FFN82} at stellar
core densities in the range $\rho Y_{e} = 10^{3} g cm^{-3}$ to
$10^{7} g cm^{-3}$. As the stellar core shifts to higher
temperatures, the FFN rates get enhanced by a factor of two. The
reasons for  enhancement in the FFN rates could be attributed to at
least two reasons.  FFN, like the large scale shell model
calculation, used Brink's hypothesis in their calculation (Brink's
hypothesis states that GT strength distribution on excited states is
\textit{identical} to that from ground state, shifted \textit{only}
by the excitation energy of the state).  FFN did not take into
account the particle emission processes from higher excited states.
As a result the parent excited states and resulting GT transitions
(using Brink's hypothesis) extended beyond particle emission
threshold energies. These states had a finite occupation probability
at high temperatures and consequently significant contribution to
total electron capture rates as can be seen from
Eq.~\ref{GrindEQ__15_}. Another reason for enhancement in the
capture rates of FFN is due to the placement of GT centroid at too
low energy in daughter nucleus (for details see
\cite{Caurier99,Langanke98,Langanke00}). In high density region the
FFN rates are bigger by roughly an order of magnitude for all
temperature ranges for reasons already mentioned.

For modeling and simulation of a successful explosion mechanism one
needs weak interaction mediated rates for chain of hundreds of
nuclei as these weak rates are used as important nuclear physics
input parameter in the simulation codes for core-collapse
supernovae. Few of such important weak rates were presented earlier
(e.g. \cite{Rahman07,Nabi07a,Nabi07b,Nabi08a,Nabi08b,Nabi09}). Work
is still in progress for improved microscopic calculation of weak
rates of remaining key iron-regime nuclei.

\section{Summary}
GT$_{+}$ transitions largely determine electron capture rates in the
stellar core. Consequently GT strength is an important ingredient
related to the complex dynamics of presupernova and supernova
explosion. The GT$_{+}$ distribution for odd-Z $^{51}$V nucleus has
been calculated within the domain of pn-QRPA theory using a
luxurious model space of $7\hbar\omega$.  The comparison of the
pn-QRPA model with high resolution (d,$^{2}$He) experiment, with
experimental results of (n, p), and with theoretical calculations of
shell model and FFN has been made. The total GT strength and the GT
centroid calculated within the domain of pn-QRPA is in good
agreement with reported high resolution (d,$^{2}$He) experimental
value. The stellar weak decay rates are sensitive to the location of
GT centroid and structure of these low-lying states in daughter
$^{51}$Ti. The small changes in the binding and excitation energies
lead to significant modifications of the predictions for the
synthesis of elements in the stellar kline. The good agreement of
the low-lying pn-QRPA calculated GT strength  with experimental
results can affect the prediction and synthesis of $^{51}$Ti as well
as the time evolution and dynamics of the collapsing supermassive
stars. FFN calculation placed the GT centroid of the GTR at 3.83 MeV
 excitation energy in daughter. The pn-QRPA  calculation yields GT
centroid at 4.20 MeV in daughter $^{51}$Ti, and is in remarkable
agreement with high resolution study of GT strength distribution
\cite{Bäumer03}, which placed centroid of the GT strength at 4.1
$\pm$ 0.4 MeV. The shell model calculations placed it at 4.34 MeV.
Within the energy range measurement of around 5 MeV, the current
pn-QRPA total B(GT) strength 0.75 is also in good agreement with the
(d, $^{2}$He) experiment's total strength of 0.9 $\pm$ 0.1.

The calculated electron capture rates on $^{51}$V in stellar matter
are in excellent agreement with the large scale shell model rates
for all temperature range. Only at high density of around $\rho
Y_{e} = 10^{11} g cm^{-3}$ do the pn-QRPA rates get suppressed by a
factor of two. The corresponding comparison with FFN rates is fair
and the pn-QRPA electron capture rates are suppressed by roughly an
order of magnitude in high density region. Brachwitz et al.
\cite{Brachwitz00} performed model calculations for Type-Ia
supernovae using FFN electron capture rates yielding overproduction
of iron group nuclei. Type-Ia supernovae are responsible for about
half of the abundances of the iron-group nuclei in the galactic
evolution \cite{Cow04}. It might be interesting for the simulators
to study how the composition of the ejecta would be effected using
the present smaller electron capture rates and to compare it with
the observed abundances using supernovae simulation codes.

Realistically speaking weak interaction mediated rates of hundreds
of nuclei are involved in the complex dynamics of supernova
explosion . Incidently, the most abundant nuclei tend to have small
weak rates as they are more stable and the most reactive nuclei tend
to be present in minor quantities. It is therefore the product of
the weak rate and nuclear abundance of a particular specie which is
more important in stellar core. We are in the process to calculate
both the nuclear abundance and the weak decay rates for nuclei,
which are considered to be important in astrophysical environment,
as part of this on-going project.

\onecolumn
%TABLES
\textbf{Table 1:} Temperature and density points for which $^{51}$V
electron capture rates are calculated.
\begin{center}
\begin{tabular}
{ccc} \\ \hline $T_{9} (K)$  & $\rho Y_{e} (g cm^{-3})$
& $\mu_{e}(MeV)$ \\
 1 & 3 & 0.046\\
  3 & 3 & 0.000\\
  10 & 3 & 0.000\\
   1 & 7 & 1.200\\
   3 & 7 & 1.021\\
 10 & 7 & 0.196\\
  1 & 11 & 23.933\\
  3 & 11 & 23.925\\
 10 & 11 & 23.832\\ \hline
\end{tabular}
\end{center}
%FIGURES

\end{document}